\documentclass[conference,letterpaper]{IEEEtran}\IEEEoverridecommandlockouts
\usepackage{fancyhdr}
\usepackage{setspace}
\usepackage{amsthm}
\usepackage{amsmath}
\usepackage{amssymb}
\usepackage{epsfig}
\usepackage{amsmath,epsfig}
\usepackage{amssymb, amsthm}
\usepackage{mathrsfs}
\usepackage{algorithm,algorithmic}
\usepackage{enumerate}
\usepackage{amsthm}
\usepackage{amsmath}
\usepackage{amssymb}
\usepackage{epsfig}

\usepackage{paralist}
\usepackage{graphicx}
\newtheorem{myth}{\bf Theorem}

\newtheorem{mydef}{\bf Definition}

\newcommand{\bx}{\boldsymbol x}

\newcommand{\mK}{\mathcal{K}}
\newcommand{\mR}{\mathcal{R}}
\newcommand{\mX}{\mathcal{X}}
\newcommand{\mY}{\mathcal{Y}}
\newcommand{\mV}{\mathcal{V}}

\newcommand{\mT}{\mathcal{T}}
\ifCLASSINFOpdf
\else
\fi
\begin{document}
%
\title{Broadcast Repair for Wireless Distributed Storage Systems
\thanks{This work was supported in part by a grant from the University Grants
Committee of the Hong Kong Special Administrative Region, China, under
Project AoE/E-02/08.}
\thanks{This work was supported in part by ARC DP150103658 and DP130102228.}
}
\author{
 \IEEEauthorblockN{Ping Hu}
 \IEEEauthorblockA{Department of Electronic Engineering\\
City University of Hong Kong\\
Email: ping.hu@my.cityu.edu.hk}

\and
 \IEEEauthorblockN{Chi Wan Sung}
 \IEEEauthorblockA{Department of Electronic Engineering\\
City University of Hong Kong\\
Email: albert.sung@cityu.edu.hk}

\and
\IEEEauthorblockN{Terence H. Chan}
\IEEEauthorblockA
{ Institute for Telecommunications Research\\ University of South Australia\\
     Email: terence.chan@unisa.edu.au}
}


%


\maketitle

\thispagestyle{fancy}
\fancyhead{}
\lhead{}
\lfoot{}
\cfoot{}
\rfoot{}
\renewcommand{\headrulewidth}{0pt}
\renewcommand{\footrulewidth}{0pt}

\begin{abstract}
In wireless distributed storage systems, storage nodes\- are
connected by wireless channels, which are broadcast in nature.
This paper exploits this unique feature to design an efficient
repair mechanism, called broadcast repair, for wireless
distributed storage systems with multiple-node failures. Since
wireless channels are typically bandwidth limited, we advocate a
new measure on repair performance called repair-transmission
bandwidth, which measures the average
number of packets transmitted by helper nodes per failed node. The fundamental
tradeoff between storage amount and repair-transmission bandwidth
is obtained. It is shown that broadcast repair outperforms cooperative repair, which is
the basic repair method for wired distributed storage systems with multiple-node failures,
in terms of storage efficiency and repair-transmission bandwidth, thus
yielding a better tradeoff curve.
\end{abstract}


%
\IEEEpeerreviewmaketitle

\section{Introduction}
In distributed storage systems (DSS), repairability is an important design issue.
Since storage node failures are common, it is important that failed nodes
can be repaired in an efficient manner.
Traditional erasure codes have high
storage efficiency, but require a large amount of data exchange (called {\em repair bandwidth})
for node repair. Dimakis \emph{et al}. show that there is a fundamental tradeoff between
storage efficiency and repair bandwidth~\cite{Dimakis}. By using \emph{information flow graph},
the repair dynamics of a DSS is modeled as a multicast network. The storage
capacity of a DSS is shown to be equal to the min-cut of the information flow graph~\cite{networkcoding}.
Furthermore, the optimal storage-repair bandwidth tradeoff can be achieved by linear network
codes with finite alphabet size even though the information flow graph is unbounded~\cite{wuyunnan_jsac}.
These fundamental results are re-examined in~\cite{itw15}.

The seminal work~\cite{Dimakis} has stimulated a lot of study
on efficient repair of failed nodes in DSS.
Most of these works focus on \emph{single-node repair}, meaning that nodes are assumed to be failed one by one and the
repair process is triggered immediately when a node failure occurs. In~\cite{Cooperative_hu}, it was observed that repair
bandwidth per failed node can be reduced if the repair is triggered only when the number of failed nodes
reaches a predetermined threshold. This mechanism is termed {\em
cooperative repair}. In~\cite{kenneth_jnl}, results on cooperative
repair are further extended to a more general scenario, and the
fundamental tradeoff
between storage amount and repair bandwidth for DSS with cooperative repair is derived.

Due to the increasing use of wireless devices and popularity of
wireless sensor networks, wireless distributed
storage systems (WDSS) has become an emerging new area.
In \cite{wireless_network_coding}~\cite{wireless_download}, more effective file retrieval methods are considered.
Concerning the repair problem, while designs for DSS can also be
applied to WDSS, it is important to understand the fundamental
difference between DSS and WDSS.
In~\cite{xiaoming14},
the transmissions
between storage nodes during repair are assumed to experience
erasures, and the fundamental storage-bandwidth tradeoff for
single-node repair in DSS with erasure channels are established,
which addresses
the issue that wireless channels are inherently unreliable.
Another basic characteristic of the wireless medium, which
distinguishes it from wired transmission, is its {\em broadcast}
nature. In~\cite{xiaoming_broadcast}, the authors studied the repair problem when parts of stored packets in nodes are lost.
They focused on one repair round and obtained the minimum transmitted packets for repair. For a special parameter setting, an exact repair code construction is proposed.
Repair under multiple repair rounds is unclear for general parameter settings.
In this paper, this broadcast nature of the communication channel in WDSS is investigated for multiple repair rounds.
To design
an efficient WDSS, the broadcast nature of the wireless medium
can be exploited during the repair process when there are more
than one failed nodes.
To reap the potential gain,
we propose {\em broadcast repair}
for WDSS with multiple node failures. A graph representation for WDSS
is constructed. By analyzing the min-cut of the graph, a bound on storage capacity is derived. Tightness of the bound is also shown. 
To quantify the benefit of broadcast repair, we compare our method with cooperative
repair with unicast transmissions and show its superiority. 

\section{System Model and Broadcast Repair}\label{sec:model}
The system model is designed to capture the broadcast
characteristic under the wireless scenario. It includes one source
node, multiple storage nodes, and multiple data collectors. Each
storage node can store $\alpha$ packets at most. Storage nodes are not directly
connected by wires. Instead, they are fully connected by a wireless broadcast medium.

At the initial stage, the source node stores a file into $n$
storage nodes such that the data collectors can
retrieve the file from any $k$ nodes. We
index these storage nodes by the set $\mathcal{N}\triangleq
\{1,2,\dots,n\}$. These $n$ storage nodes are not reliable and can fail at times becoming inactive. When the number of failed nodes is
accumulated up to a threshold $r$, the repair
process is triggered. We call this process one round of repair. During each
repair round, $r$ new nodes, called \emph{newcomers}, will join
the system. Then $d\geq k$ active storage nodes will broadcast
packets to the newcomers. Each of these $d$ nodes, called
\emph{helper node}, will broadcast $\beta$ packets. We assume that $r \leq n - d$ so that there are always $d$ active nodes in the system for repair.
When a helper node broadcasts a packet, we assume that each
newcomer receives the packet successfully without error.
Besides, we also assume the helper nodes use orthogonal channels to
transmit their packets so that there is no interference between
their transmissions. We consider $T$ rounds of repair in total. After each repair round, there are $r$ newcomers,
which replace the $r$ failed nodes in this round. We index the
newcomers after the $s$-th repair round by
$\mathcal{R}_s\triangleq\{n+(s-1)r+1,\dots,n+sr\}$. The set of
helper nodes for these newcomers are denoted by $\mathcal{H}_s$.

Any data collector can join the system after the initialization
stage or after any repair round. It can connect to any $k$
active nodes to retrieve all data stored in the node. Denote the data collector which joins after the $s$-th round of repair and connects to a set $\mK$ of $k$ active nodes by $\mathsf{DC}_{s,\mK}$.
Since we have to ensure that the file can always be retrieved, we consider all possible arrivals (in terms of $s$) and connections (in terms of $\mK$) of a data collector. We denote the collection of all possible data collectors by $\Omega$.

The above system is called a WDSS with parameters
$(n,k,d,r,\alpha,\beta,T)$. The repair process described
above is called \emph{broadcast repair}. An \emph{instance} of a
WDSS is determined by the failure patterns, newcomers, and the collection of helper sets.


In the literature of DSS, the total number of packets downloaded
by a newcomer so as to repair a failed node is called {\em repair
bandwidth}~\cite{Dimakis}. It is one of the key performance
metrics in DSS, since it reflects the amount of network traffic
required in the repair process. The same concept can also be applied to
multiple node failures with cooperative repair
processes~\cite{Cooperative_hu}. In a wireless environment,
however, repair bandwidth is not an accurate measure on network
traffic, especially when there are multiple node failures. To see
this, consider the double-failure case where two newcomers receive
the same packet from the wireless broadcast of a helper node.
While this broadcast packet would be counted twice in calculating
repair bandwidth, this packet was broadcasted only once. To better reflect the use of frequency spectrum
in a wireless environment, we introduce a new performance metric
named \emph{repair-transmission bandwidth}:

\begin{mydef}
 The repair-transmission bandwidth, $\tau$, is defined as the average number of packets the helper nodes transmitted per newcomer.
\end{mydef}

If all the packet transmissions are in unicast mode, then
repair-transmission bandwidth is equal to repair bandwidth, since
the total number of packets transmitted by helper nodes is equal
to the total number of packets received by newcomers. They are
different, however, when broadcast transmissions are allowed. For
the WDSS model described above, we have
$$
\tau=\frac{d\beta}{r}.
$$
Given a requirement on storage capacity $C_{\text{storage}}$,
there is a trade-off between the per-node storage capacity,
$\alpha$, and the repair-transmission bandwidth, $\tau$.


\section{Graph Representation}\label{sec:graph}

The network of WDSS is represented by a directed acyclic graph $G=(\mathcal{V},\mathcal{E})$, where $\mathcal{V}$ is the vertex set and $\mathcal{E}$ is the edge set.
Each edge $e(i,j)\in \mathcal{E}$ is associated with a
parameter $u_{ij}$, which is the capacity of the edge.

The graph includes one source vertex $\mathsf{S}$, multiple
storage nodes (including failed ones and newcomers), and multiple
data collectors $\mathsf{DC}_{s,\mathcal{K}}$'s. Each
storage node $j$ is represented by two vertices, \emph{in-vertex} $\mathsf{In}_j$,
\emph{out-vertex} $\mathsf{Out}_j$, and a directed edge
$\mathsf{In}_j\to\mathsf{Out}_j$ with parameter $\alpha$.
In this paper, the terms ``node'' and ``vertex'' have different
meanings. A node refers to a storage device in the WDSS while a vertex is an abstract entity of
the graph.

In the initialization stage where data is first stored at the storage nodes, the source vertex $\mathsf{S}$
transmits packets to the storage nodes and then becomes inactive. This is
modeled by adding the edges $\mathsf{S}\to\mathsf{In}_j$, for all
$j \in \mathcal{N}$, with capacity $\infty$.

\begin{figure}
 \centering
 \includegraphics[width=3.7in,angle=0]{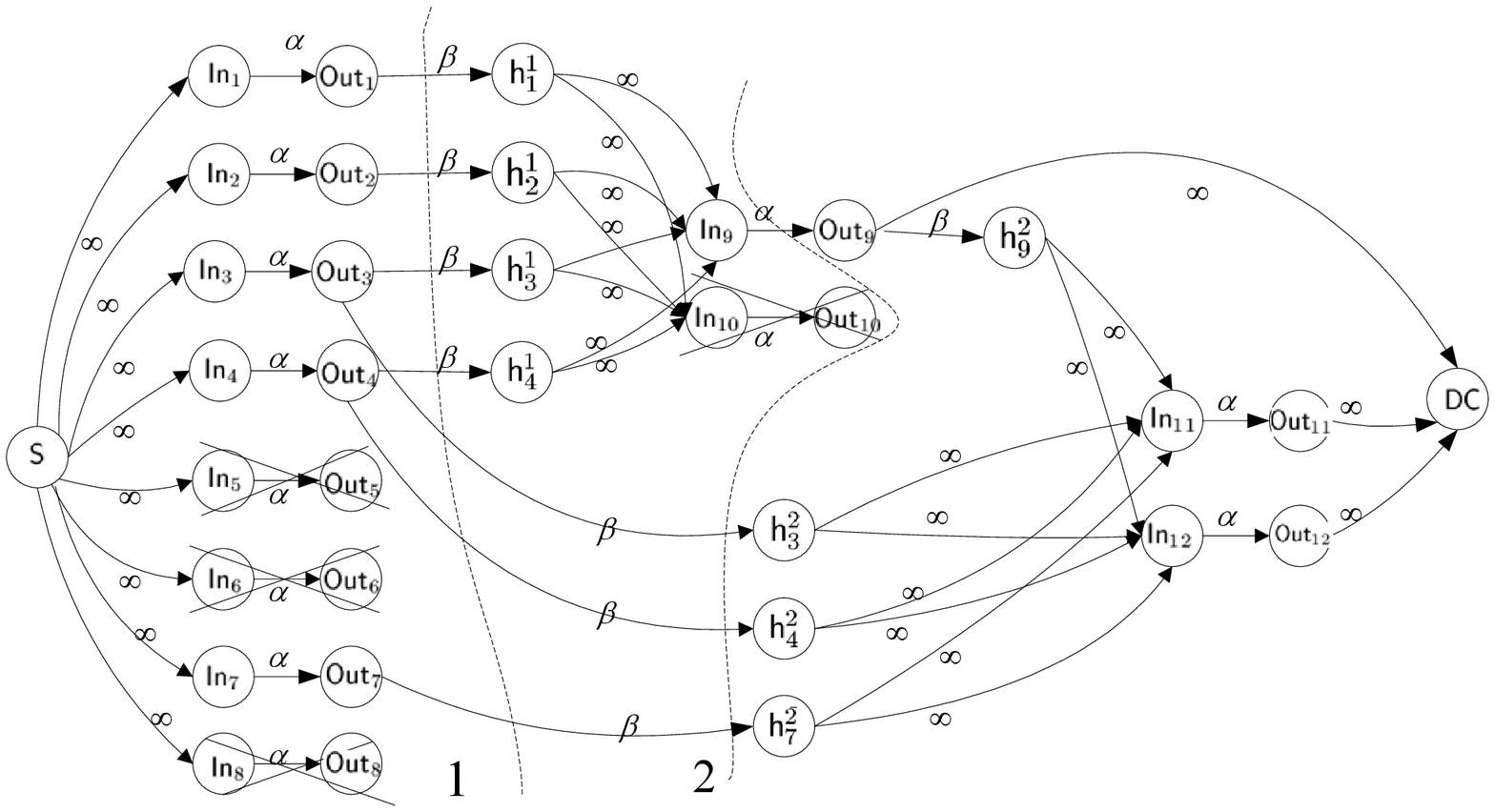}
 \caption{An example for cut-capacity in broadcast repair.}~\label{fig:define_example}
\end{figure}

In the first repair round (i.e., $s=1$), node $i\in\mathcal{H}_1$ broadcasts $\beta$ packets to newcomer node $j\in\mathcal{R}_1$, which is again modeled by two vertices $\mathsf{In}_j$, $\mathsf{Out}_j$, and a directed edge $\mathsf{In}_j\to\mathsf{Out}_j$ with parameter $\alpha$.
Note that $\mathcal{R}_1$ and $\mathcal{N}$ are disjoint, meaning that a newcomer has a new index, which is different from the index of the failed node being replaced by that newcomer. For helper node $i\in\mathcal{H}_1$, we add an \emph{auxiliary vertex}, say $\mathsf{h}_i^1$, to which $\mathsf{Out}_i$ is connected by an edge with capacity $\beta$. Edges with capacity $\infty$ are added from vertex $\mathsf{h}_i^1$ to $\mathsf{In}_j$ of every newcomer $j \in \mathcal{R}_1$.
The vertex $\mathsf{h}_i^1$ is used to model the broadcast feature of the wireless channel. Subsequent repair rounds are modeled in the same way. Consider the example shown in Fig.~\ref{fig:define_example}. The corresponding WDSS has parameters $n=8, k=3, d=4,$ $r=2$ and $T=2$. In this example, nodes $5$ and $6$ failed in the first repair round, and we have $\mathcal{R}_1=\{9,10\}$ and $\mathcal{H}_1=\{1,2,3,4\}$.
Nodes $8$ and $10$ failed in the second repair round, and we have $\mathcal{R}_2=\{11,12\}$ and $\mathcal{H}_2=\{9,3,4,7\}$.


To model the file retrieval process, after each repair round~$s$ and
for each possible choice of $\mK$, we add a data collector $\mathsf{DC}_{s,\mK}$.
Furthermore, a directed edge from each out-vertex of a node in $\mK$ to
$\mathsf{DC}_{s,\mK}$ with capacity $\infty$ is
added. In Fig.~\ref{fig:define_example}, we show only one data collector, namely, $\mathsf{DC}_{2,\{9,11,12\}}$,
for simplicity.

An $\mathsf{S}$-$\mathsf{DC}$ {\em cut} $\mX$ is a subset of $\mV$ such that $\mathsf{S}\in \mathcal{X}$, $\mathsf{DC} \in \mathcal{Y} \triangleq \mV \setminus \mX$ and there is at least one edge from $\mX$ to $\mY$. The {\em cut-set} of a cut $\mX$ is $\{(i,j) \in \mathcal{E} : i\in \mX, j \in \mY\}$.
The {\em cut-capacity} of $\mX$ is defined as:
\begin{equation}
C(\mX)\triangleq\sum_{(i,j) \in \mathcal{E},i\in\mX,j\in\mY} u_{ij}.\label{eq:def_cut}
\end{equation}
Two examples of $\mathsf{S}$-$\mathsf{DC}_{2,\{9,11,12\}}$ cuts are denoted in
Fig.~\ref{fig:define_example} by left sides of dashed lines.
For line $1$, the cut-capacity is $7\beta$, while for line
$2$, the cut-capacity is $ \alpha+3\beta.$

%

\section{Storage Capacity}\label{sec:capacity}


According to~\cite{networkcoding}, the capacity of the single source multicast network is given by the minimum value of the cut-capacity between the source node and any
of the destinations. Therefore, the storage capacity of a particular WDSS instance $I$ is given by
\begin{align}
C'(I) = \min_{\mathsf{DC}} \min_{\mX:\mathsf{S}-\mathsf{DC}\text{~cut}}C(\mX),\label{eq:def_capa_instance}
\end{align}
where the first minimum is taken over all legitimate choices of $\mathsf{DC}$ under the instance $I$.
The storage capacity, $C_{\text{storage}}$, of a WDSS can be
obtained by minimizing $C'(I)$ over all its possible instances, i.e.,
\begin{align}
C_{\text{storage}}=\min_{I}C'(I).\label{eq:def_capa}
\end{align}

Consider an arbitrary instance of the WDSS.
Regard the initialization stage as round 0 and let $\mathcal{V}_0\triangleq \mathcal{N}$. For rounds $s\in\mT\triangleq\{1,2,\ldots,T\}$, let $\mathcal{A}_s \triangleq \{\mathsf{h}_i^s : i\in\mathcal{H}_s\}$ be the set of auxiliary vertices in round~$s$, and $\mathcal{V}_s\triangleq \mathcal{A}_s \cup \{\mathsf{In}_j, \mathsf{Out}_j : j \in \mR_s\}$ be the set of all vertices in round~$s$. Then $\mathcal{V}_0 \cup \mathcal{V}_1\cup \cdots \cup \mathcal{V}_T$ contains all the vertices except the source and the destinations in the graph.

To obtain the cut-capacity of an arbitrary $\mathsf{S}-\mathsf{DC}$ cut $\mX^{\prime}$, we examine the in-edges of all the vertices in $\mathcal{V}_0\cup\mathcal{V}_1\cup \cdots \cup\mathcal{V}_T$, and express the cut-capacity as a sum of $T+1$ terms:
\begin{equation}
C(\mX^{\prime})= \sum_{0\leq s\leq T}C_{\vartriangle,s}(\mX^{\prime}), \label{eq:sum_contri}
\end{equation}
where $$C_{\vartriangle,s}(\mX^{\prime})  \triangleq \sum_{i\in\mX^{\prime}, j\in \mathcal{V}_s\cap \overline{\mX^{\prime}}}u_{ij}$$ is called the \emph{cut-capacity contribution} of the vertices in $\mathcal{V}_s$. When there is no ambiguity, we may simply write it as $C_{\vartriangle,s}$. For example, in Fig.~\ref{fig:define_example}, the cut denoted by left side of line~2 has cut-capacity equal to $C_{\vartriangle,0}+C_{\vartriangle,1}+C_{\vartriangle,2} = 0 + \alpha + 3\beta$.


Now we investigate $C_{\vartriangle,s}(\mX^{\prime})$ for $0\leq s\leq T$. First, consider the case where $s=0$. Since there is no auxiliary vertex in repair round~0, we have
$
C_{\vartriangle,0}=x_0\alpha,
$
where $x_0$ is the number of storage nodes in round~0 such that its in-vertex is in $\mX^{\prime}$ and out-vertex is in $\overline{\mX^{\prime}}$.



\begin{figure}
 \centering
 \includegraphics[width=1.8in,angle=0]{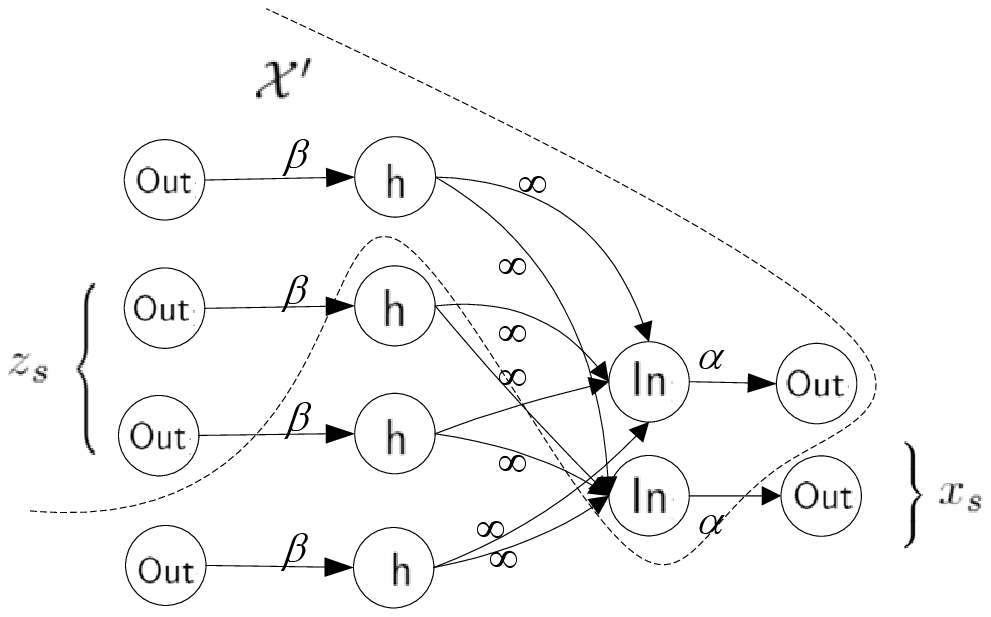}
 \caption{An illustration of case where $s\in \mT_1^{\prime}$.}~\label{fig:analysis_T1}
\end{figure}

Second, consider the case where $s\in \mT_1^{\prime}$, where $\mT_1' \triangleq \{s \in \mT : \mathcal{A}_s \cap \mX' \neq \emptyset\}$. In other words, $s \in \mT_1'$ if there exists at least one $\mathsf{h}_i^s$ in $\mX^{\prime}$. We investigate the three classes of vertices in $\mathcal{V}_s$, i.e., auxiliary vertices, in-vertices, and out-vertices, one by one. For the auxiliary vertices, denote the number of $\mathsf{h}_i^s$'s such that it is in $\overline{\mX^{\prime}}$ and its parent vertex $\mathsf{Out}_i$ is in $\mX^{\prime}$ by $z_s$. For the in-vertices, we only need to consider the case where all of them are in $\mX^{\prime}$, for otherwise, the cut-capacity contribution would be infinite as all its in-edges have infinite capacity and at least one of its parent vertex is in $\mX^{\prime}$. For the out-vertices, denote the number of them in $\overline{\mX^{\prime}}$ by $x_s$. Then we have
$
C_{\vartriangle,s}=x_s\alpha+z_s\beta.
$
An illustration of this case is shown in Fig.~\ref{fig:analysis_T1}

\begin{figure}
 \centering
 \includegraphics[width=1.8in,angle=0]{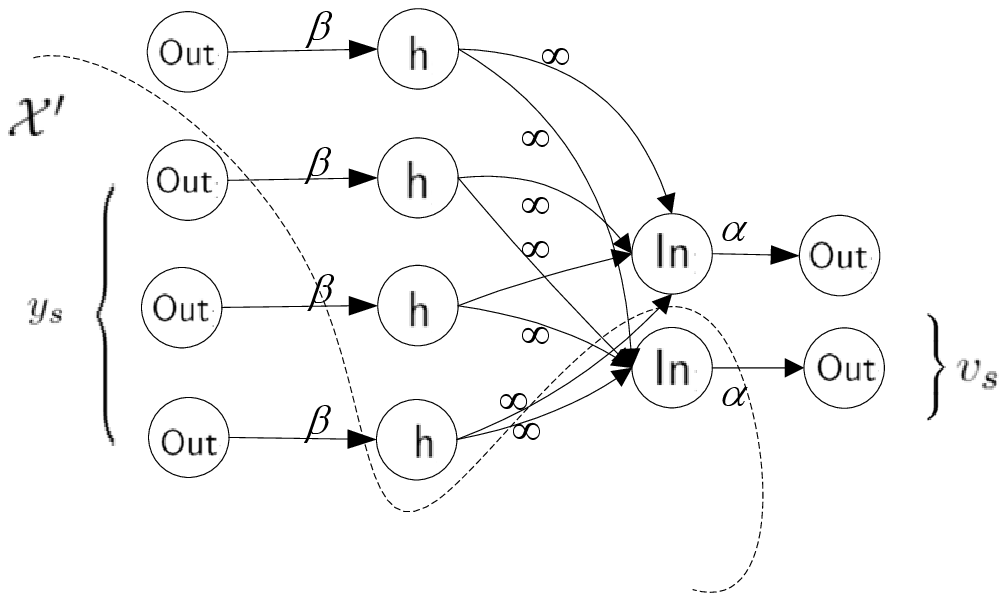}
 \caption{An illustration of case where $s\in \mT_2^{\prime}$.}~\label{fig:analysis_T2}
\end{figure}

Third, consider the case where $s\in \mT_2^{\prime}\triangleq\mT\setminus \mT_1 ^{\prime}$. By definition of $\mT_2^{\prime}$, all $\mathsf{h}_i^s$'s are in $\overline{\mX^{\prime}}$.
For the auxiliary vertices, denote the number of $\mathsf{h}_i^s$'s such that its parent vertex $\mathsf{Out}_i$ is in $\mX^{\prime}$ by
$y_s$. For all the in-vertices, since their parent vertices are all in $\overline{\mX^{\prime}}$,
their cut-capacity contribution is zero, no matter they are in $\mX^{\prime}$ or $\overline{\mX^{\prime}}$. The cut-capacity contribution of all the in-vertices is 0. For the out-vertices, denote the number of them such that $\mathsf{Out}_j\in \overline{\mX^{\prime}}$ and its parent vertex $\mathsf{In}_j\in \mX^{\prime}$ by $v_s$.
we have
$
C_{\vartriangle,s}=v_s\alpha+y_s\beta.
$
An illustration of this case is shown in Fig.~\ref{fig:analysis_T2}

Combining the above three cases and according to~\eqref{eq:sum_contri}, we have
\begin{align}
C(\mX^{\prime})=x_0\alpha+\sum_{s\in \mT_1^{\prime}}\left(x_s\alpha+z_s\beta\right)+\sum_{s\in \mT_2^{\prime}}(v_s\alpha+y_s\beta)\notag.
\end{align}

Now consider a special cut $\mX$, which is constructed from $\mX^{\prime}$ as follows. Initially, let $\mX$ be the same as $\mX^{\prime}$. For $s\in \mT_1^{\prime}$, move all $\mathsf{h}_i^s$'s into $\mX$, and then $z_s$ becomes zero.
Furthermore, since $\mathsf{h}_i^s$'s child vertices are all in round $s$, moving all $\mathsf{h}_i$'s into $\mX$ will not affect the cut-capacity contribution of other rounds. For $s\in \mT_2^{\prime}$, move all $\mathsf{In}_j$'s into $\overline{\mX}$, and $v_s$ becomes zero. Again, since $\mathsf{In}_j$'s child vertex $\mathsf{Out}_j$ is in the same round, moving $\mathsf{In}_j$ will not affect the cut-capacity contribution of other rounds.
We have
\begin{align}
C(\mX^{\prime})\geq C(\mX)=x_0\alpha+\sum_{s\in \mT_1}x_s\alpha+\sum_{s\in \mT_2}y_s\beta,\label{eq:bound_x'}
\end{align}
where $\mT_1$ is the index set of repair rounds whose auxiliary vertices are all in $\mX$, and $\mT_2$ is the index set of repair rounds whose auxiliary vertices are all in $\overline{\mX}$. Note that $\mT_1 \cup \mT_2 = \mT$. 
Since $x_s$ is the number of vertices such that it is in $\overline{\mX}$ and its parent vertex is in $\mX$, let $x_s=0$ for $s\in \mT_2$.

For {\em any} round $s$, let the number of out-vertices in $\overline{\mX}$ be $m_s$. By definition, we have
$m_s = x_s,   \; \; \text{if } s\in \mT_1\cup\{0\}$ and
\begin{align}
&0\leq m_s \leq r, \;\; \text{if } s\in \mT_2. \label{m_s_in_T2}
\end{align}
Furthermore, we define
\begin{equation}
m_s^*=\left\{\begin{array}{ll} x_s, & s\in \mT_1\cup\{0\} \\ r, & s\in \mT_2.\end{array}\right.\label{eq:def_m^*}
\end{equation}
The following result is established by finding bounds for $x_s$ and $y_s$.

\begin{myth}\label{thm:bound}
The storage capacity of WDSS$(n,k,d,r,\alpha,\beta,T)$ $C_{\text{storage}}$ is lower bounded by $C_{LB}(T)\triangleq $ \begin{align}
\min_{\bx,\mT_1}\left\{x_0\alpha+\sum_{s\in \mT_1}x_s\alpha+\sum_{s\in \mT\setminus\mT_1}\max(0,d-\sum_{i=0}^{s-1}m_i^*)\beta\right\},\label{eq:bound}
\end{align}
where the minimization is taken over $\mT_1\subseteq \mT = \{1, 2, \ldots , T\}$ and
\begin{align}
  &0\leq x_0\leq n,\label{eq:con_x0_real}\\
  &0 \leq x_s\leq r, \text{ for } s\in \mathcal{T}_1, \label{eq:xs_range} \\
 k\leq \sum_{s\in\mT_1\cup\{0\}} x_s & + (|\mT|-|\mT_1|)r\leq k+r. \label{eq:sum_xs}
\end{align}
\end{myth}

\begin{proof}
Since any arbitrary data collector $\mathsf{DC}_{s,\mK}$ is able to connect to $k$ out-vertices through links with infinite capacity, for any cut with finite cut-capacity, we must have
\begin{equation}
  \sum_{s\in\mT_1\cup\{0\}}x_s+\sum_{s\in\mT_2}m_s\geq k. \label{eq:collector}
\end{equation}
It is clear that $x_s$ must also satisfy~\eqref{eq:con_x0_real} and~\eqref{eq:xs_range}.

There are $d$ helper nodes for newcomers in round~$s$ and at most $\sum_{i=0}^{s-1}m_i$ of them have their out-vertices in $\overline{\mX}$. Therefore, we have
\begin{align*}
y_s\geq \max(0,d-\sum_{i=0}^{s-1}m_i),\text { for } s\in \mT_2.
\end{align*}
By \eqref{eq:bound_x'}, the cut-capacity of a given cut is bounded below by
$$
x_0\alpha+\sum_{s\in \mT_1}x_s\alpha+\sum_{s\in \mT_2}\max(0,d-\sum_{i=0}^{s-1}m_i)\beta.
$$
Note that the above expression is a monotonic decreasing function of each $m_i$. If \eqref{eq:collector} is originally satisfied, increasing the value of each $m_i$ will not violate it. By \eqref{m_s_in_T2}, the above expression is minimized when $m_s^* = r$ for $s \in \mT_2$. Then~\eqref{eq:collector} becomes
\begin{equation}
\sum_{s\in\mT_1\cup\{0\}} x_s  + \sum_{s\in\mT_2} r \geq k.\label{eq:sum_xs_original}
\end{equation}


Let $(\bx^*,\mT_1^*)$ achieves the minimum of~\eqref{eq:bound}, where $\bx^* \triangleq(x_0^*, x_1^*, \ldots, x_T^*)$. The minimum value of~\eqref{eq:bound} is thus equal to
\begin{align}
& \sum_{s\in \mT_{1}\cup\{0\}}x_s^* \alpha+\sum_{s\in \mT_{2}}\max(0,d-\sum_{i\in \mT_1\cup\{0\},\atop i<s}x_i^*-\sum_{i\in \mT_2,i<s}r)\beta.\label{eq:assume_optimal_value}
\end{align}
Suppose to the contrary that
\begin{equation}
\sum_{s\in\mT_1\cup\{0\}}x_s^* + \sum_{s\in\mT_2}r>k+r.\label{eq:assumption}
\end{equation}
We claim that we can always find another feasible solution which achieves a lower objective function value than~\eqref{eq:assume_optimal_value}. To see this, consider the last repair round $s' \in \mT$ which has strictly positive value of $m_{s'}^*$. If $m^*_T > 0$, then $s' = T$. Otherwise, we have $m_{s'}^* > 0$ and $m_s^* = 0$ for all $s > s'$.
By definition of $m^*_s$, we have $s\in\mT_1$ for all $s > s'$.
If $s'\in \mT_1$, we can set $x_{s'}^*=0$ to strictly reduce the value of the expression in~\eqref{eq:assume_optimal_value}. Since $x_{s'}^*\leq r$, the new setting will not violate constraint~\eqref{eq:sum_xs_original}. Therefore,~\eqref{eq:assumption} cannot hold. If $s'\in \mT_2$, we move $s'$ from $\mT_2$ to $\mT_1$, and set $m_{s'}^*=0$. The value in~\eqref{eq:assume_optimal_value} will not be increased while constraint~\eqref{eq:sum_xs_original} will still be satisfied due to the assumption in~\eqref{eq:assumption}. We then repeat the above argument and find another new index $s'$. Due to constraint~\eqref{eq:sum_xs_original}, $\bx^*$ cannot be the zero vector, which leads to a contradiction. Hence,~\eqref{eq:assumption} does not hold and we must have~\eqref{eq:sum_xs}.
\end{proof}

%


In the following theorem, we show the tightness of the lower bound when $T$ is finite.
\begin{myth}\label{thm:capacity_achieve}
When $T$ is finite, the lower bound in Theorem~\ref{thm:bound} is tight when $n\geq k+2r$.
\end{myth}

\begin{proof}
Let $(\bx^*,\mT_1^*)$ be an optimal solution to the minimization in Theorem~\ref{thm:bound}.
We prove that the bound is tight by constructing an instance $I^*$ with a
$\mathsf{DC}^*$ and a cut $\mX^*$ such that the cut-capacity
$C(\mX^*)$ is exactly $C_{LB}(T)$. 

The instance $I^*$ is constructed as follows. First, in stage 0, choose any $r$ nodes in $\mathcal{N}$ and let them fail. For stage $s\in\mT_1^*$, choose any $r-x_s^*$ nodes in $\mR_s$ and any $x_s^*$ active nodes in $\mathcal{N}$ and let them fail right before stage $s+1$. For stage $s\in\mT_2^*$, choose any $r$ active nodes in $\mathcal{N}$ and let them fail right before stage $s+1$. We can always find such a failure pattern since there are $r$ nodes in $\mR_s$ for every $s$, and the accumulated number of failed nodes in $\mathcal{N}$ is
$$
r+\sum_{s\in\mT^*_1}x_s^*+\sum_{s\in\mT_2^*}r\leq k+2r\leq n,
$$
where the first inequality follows from~\eqref{eq:sum_xs} and the second inequality follows our assumption in Theorem~\ref{thm:capacity_achieve}. Select any $x_0^*$ active nodes from $\mathcal{N}$ and denote them by $\mathcal{M}_0$. Denote the active nodes in $\mR_s$ by $\mathcal{M}_s$. In other words, for $s\in\mT_1^*\cup \{0\}$, $\mathcal{M}_s\subseteq\mR_s$ and $|\mathcal{M}_s|=x_s^*$; for $s\in\mT_2^*$, $\mathcal{M}_s=\mR_s$.

Next, we specify the helper nodes for each repair round. The helper nodes for repair round $i$, for $i = 1, 2, \ldots, s$, are chosen first from $\mathcal{M}_0$, then from $\mathcal{M}_1$, and so on, until $d$ helper nodes are chosen. If $\sum_{i=0}^{s-1}|\mathcal{M}_i|<d$, the remaining helper nodes are chosen arbitrarily from the active nodes in $\mathcal{N}$. The existence of such a helper pattern is validated by
$$
n-r-|\mathcal{M}_1|-\dots-|\mathcal{M}_{s-1}|\geq d-|\mathcal{M}_0|-\dots-|\mathcal{M}_{s-1}|,
$$
where the left side is the number of active nodes in $\mathcal{N}$ after stage $s-1$, and the right side is the number of required helper nodes in $\mathcal{N}$. The inequality holds because $n-r\geq d$.


Finally, consider $\mathsf{DC}_{T,\mK^*}$, which comes after the
repair round $T$ and connects to $\mK^* \subseteq \mathcal{M}_0 \cup \mathcal{M}_1 \cup \dots \cup \mathcal{M}_T$.
Note that there is such a DC, since according to~\eqref{eq:sum_xs}, $|\mathcal{M}_0\cup\mathcal{M}_1 \cup \dots \cup \mathcal{M}_T|\geq k$.

The cut $\mX^*$ is constructed as follows: For $s\in \{0\}\cup \mT_1^*$, put $\mathsf{Out}_i$, for $i\in\mathcal{M}_s$, into $\overline{\mX^*}$, and all the remaining vertices in round~$s$ into $\mX^*$.
Vertices in these repair rounds contribute $x_0^*\alpha+\sum_{s\in \mT_{1}^*}x_s^*\alpha$ to the cut-capacity.
For $s\in \mT_2^*$, put all vertices in round~$s$ into $\overline{\mX^*}$. Vertices in these repair rounds contribute $\sum_{s\in \mT_{2}^*}\max \Big \{ 0,d-x_0^*-\sum_{i\in \mT_1^*, i < s}x_i^*-\sum_{i\in \mT_2^*, i<s}r \Big \}\beta$ to the cut-capacity.
Summing up the cut-capacity contribution of all the vertices, we get $C_{LB}(T)$, 
showing that the bound in Theorem~\ref{thm:bound} is tight.
\end{proof}


Furthermore, the following result significantly reduces the dimension of the minimization problem. The proof is omitted due to space limitation.
\begin{myth}\label{thm:bound_irrelevent_S}
When $T\geq k+r$, we have $C_{LB}(T) = C_{LB}(k+r)$.
\end{myth}

%

\section{Comparison with Cooperative Repair}\label{subsec:compare}
We compare broadcast repair with cooperative repair when $T\geq k+r$ and $k=ur$, where $u$ is an integer larger than 1. Both repair process is triggered after the number of failed storage nodes accumulate to $r$. Consider the two points, \emph{minimum storage} (MS) point, which corresponds to the best storage efficiency, and the \emph{minimum repair-transmission bandwidth} (MT) point, which corresponds to the minimum repair-transmission
bandwidth on the trade-off curve between repair-transmission bandwidth and storage (see Fig.4 shown in next page for example).
In cooperative repair, the repair-transmission bandwidth is equal to the repair bandwidth. According to~\cite{kenneth_jnl}, the MS point and the MT point are
$
(\tau_{\text{MSC}},\alpha_{\text{MSC}})=(\frac{d+r-1}{k(d+r-k)},\frac{1}{k}),
\text{ and }
(\tau_{\text{MTC}},\alpha_{\text{MTC}})=\frac{2d+r-1}{k(2d+r-k)}(1,1),
$
respectively. We can also derive the MS point and
the MT point in broadcast repair, which are:
$
(\tau_{\text{MSB}},\alpha_{\text{MSB}})=\big(\frac{d}{k(d+r-k)},\frac{1}{k}\big),
\text { and }
(\tau_{\text{MTB}},\alpha_{\text{MTB}})=\frac{2d}{k(2d+r-k)}(1,1),
$
respectively. Since $r > 1$, broadcast repair outperforms cooperative
repair at these two points.

In Fig.4 (shown in next page), we plot the tradeoff curves of the two repair schemes with
parameters $C_{\text{Storage}}=1, d=9$, $k=4$ and $r=2$. As a benchmark, we also plot
the single-node repair, in which the repair is triggered whenever there is
a single node failure. As reported in~\cite{kenneth_jnl}, cooperative repair performs
better than single-node repair due to the benefit of node cooperation.
However, it performs worse than broadcast repair, since it does not exploit
the broadcast nature of the wireless medium.


%

\bibliographystyle{IEEEtran}

\bibliography{zigzag}
\end{document}